\documentclass[fleqn,10pt]{wlscirep}
\usepackage{tabularx}
\usepackage{graphicx}
\graphicspath{{pict/}{pic/}{}}

\title{Observation of nonlinear thermal optical dynamics in a chalcogenide nanobeam cavity}

\author[1,2,*]{Yue Sun}
\author[3]{Thomas P. White}
\author[2]{Duk-Yong Choi}
\author[1]{Andrey A. Sukhorukov}
\affil[1]{Nonlinear Physics Centre, Research School of Physics and Engineering, The Australian National University, Acton, ACT 2601 Australia}
\affil[2]{Laser Physics Centre, Research School of Physics and Engineering, The Australian National University, Acton, ACT 2601 Australia}
\affil[3]{Centre for Sustainable Energy Systems, Research School of Engineering, Australian National University, Acton, ACT 2601, Australia}
\affil[*]{yue.s@anu.edu.au}

\begin{abstract}
We present a theoretical and experimental analysis of nonlinear thermo-optic effects in suspended chalcogenide glass nano-beam cavities. We measure the power-dependent resonance peaks and characterise the dynamic nonlinear thermo-optic response of the cavity under modulated light input.
Several distinct nonlinear characteristics are identified, including a modified spectral response containing periodic fringes, a critical wavelength jump and saturated time-delay for modulation frequency faster than the thermal characteristic time.
We reveal that the coupling to a parasitic Fabry-P\'{e}rot cavity enables isolated thermal equilibrium states resulting in the discontinuous thermo-optic critical point.
\end{abstract}
\begin{document}

\flushbottom
\maketitle
%
%
\thispagestyle{empty}


\section*{Introduction}

Nanobeam cavities can achieve high quality factor $Q$, low mode volume $V$ and high on-resonance transmission~\cite{Quan:APL_203102:2010, Quan:OE_18529:2011}, making them an ideal platform for enhancing light-matter interactions including nonlinear optical effects. High $Q/V$ nanobeam cavities are very susceptible to thermal  nonlinearities~\cite{Carmon:OE_4742:2004} due to the high optical power density in the cavity at resonance. Accordingly, the thermo-optic bistability power threshold is usually very low~\cite{Haret:OE_21108:2009}, which can be beneficial to thermo-optic switching but detrimental to other, weaker nonlinear effects. A detailed understanding of the temporal and spectral contributions of thermal nonlinearities is essential both for maximising the benefit or minimising the hindrance of thermo-optic effects, and managing thermal properties in a tailorable manner.


Recent studies have shown the possibility to engineer the thermal properties of photonic crystals~\cite{Maldovan:Nature_209:2013} and in turn thermo-optic effects~\cite{Song:OE_4235:2013}. In bulk materials, the thermo-optic response is considered as a robust but very slow effect.
However, it was reported that the thermo-optic response time in silicon nanobeam cavities on silica substrates can be as fast as $0.5 \mu$s~\cite{Haret:OE_21108:2009} because the effective heat capacity can be very low due to the ultra-small optical mode volume. The potential for tuning the effective thermal properties of nanobeam cavities could thus provide a path towards even faster thermo-optic responses.

Here, we focus on chalcogenide glass material and demonstrate a nanobeam cavity made of $Ge_{11.5}As_{24}Se_{64.5}$ with high $Q/V$ and short thermal relaxation time. Chalcogenide glasses recently have emerged as a promising platform for integrated photonics devices due to their large Kerr nonlinearity and negligible free carrier effects~\cite{Eggleton:NatPhoto_141:2011}. In the present work, we show that their high refractive index allows the high $Q/V$ cavity and their low thermal conductivity enables fast thermal response.

We present in this paper a comprehensive experimental and theoretical study of thermal nonlinearity in suspended $Ge_{11.5}As_{24}Se_{64.5}$ nanobeam cavities. It is found that coupling between the nanobeam cavitiy and a Fabry-P\'{e}rot (FP) cavity modifies the thermo-optic effects and enables isolated thermal equilibria. We also measure the dynamical response and observe the thermal nonlinearity induced harmonic generation and time delay under modulated light input at frequency up to $2$~MHz.

\section*{Results}

\begin{figure}[h!]
   \includegraphics[width=0.98\columnwidth]{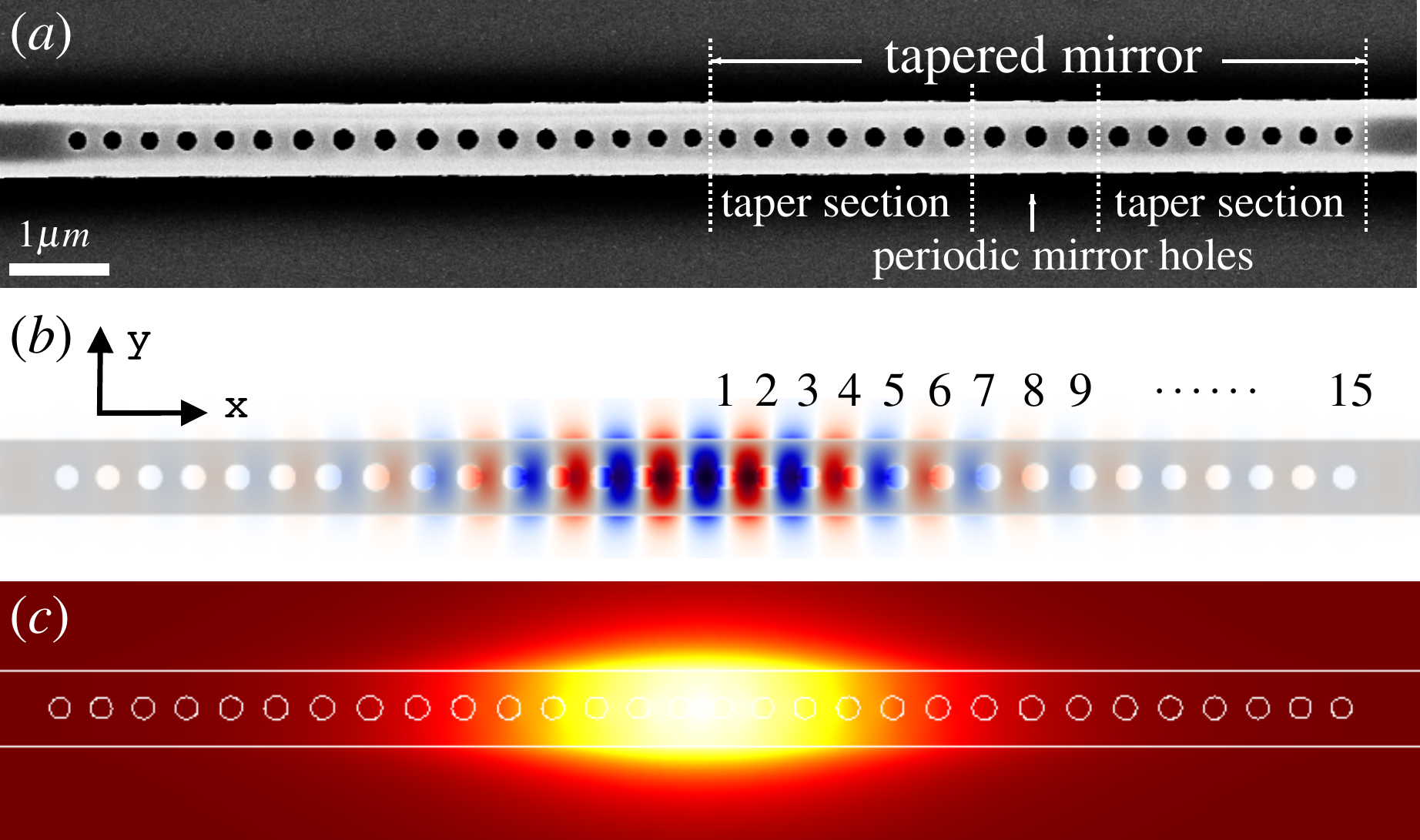}
   \caption{ (a)~Scanning electron microscope image of the fabricated chalcogenide glass nano-beam cavity. (b, c)~Simulation result of (b) $E_y$ field profile of the first order cavity mode and (c) temperature distribution of the quasi-stationary thermal state in the $z=0$ plane. The grey shading in (b) represents the chalcogenide glass cavity and the white lines in (c) corresponds to the glass-air boundary.}
   \label{fig:strsim}
\end{figure}

We designed, fabricated and measured a suspended nano-beam cavity as shown in Fig.~\ref{fig:strsim}. The waveguide and cavity were formed in a $300$ nm thick Ge$_{11.5}$As$_{24}$Se$_{64.5}$ film that was thermally-evaporated onto a SiO$_2$/Si substrate coated with $100$ nm of the polymer SU-8. The nanobeam was patterned in the chalcogenide layer using electron beam lithography followed by plasma etching. The very central $20$ $\mu$m length of the nanobeam containing the cavity was then released from the substrate by using oxygen plasma etching to remove the SU-8 sacrificial layer.  The fabricated nanobeam cavity has a width of $750$ nm, a height of $300$ nm and an air spacing between the beam and the silica substrate of $100$ nm. The two ends of the nanobeam extend to the edge of the substrate to provide access waveguides for coupling light in and out of the cavity.

\begin{table}[htb]
\caption {Air holes' radius $r$ and position $p$ in Fig.~\ref{fig:strsim}(b)}
\resizebox{\hsize}{!}{$
\begin{tabularx}{1.05\columnwidth}{|c||c|c|c|c|c|c|c|c|c|c|c|c|c|c|c|}
\hline
No. & 1 & 2 & 3 & 4 & 5 & 6 & 7 & 8 & 9 & 10 & 11 & 12 & 13 & 14 & 15\\ \hline
r [nm] & 102 & 105 & 107 & 110 & 112 & 115 & 117 & 120 &117 & 115 & 112& 110 & 107 & 105 & 102\\ \hline
p [nm] & 201 & 601 & 1011 & 1431 1& 1861 & 2301 & 2751 & 3211& 3671 & 4121 & 4561& 4991 & 5411 & 5821&6221 \\ \hline
\end{tabularx}$}
\label{tab:HoleSize}
\end{table}

The optical cavity is created by the row of air holes etched along the beam. The arrangement of air holes forms two identical tapered one-dimensional photonic crystal mirrors which are positioned end to end. The radius and position of each hole in the tapered mirror shown in Fig.~\ref{fig:strsim}(b) are listed in Tab.\ref{tab:HoleSize}. We design the tapered mirrors by minimising the radiation loss and optimising the on-resonance transmission through matching the Bloch mode index, using a similar approach to that reported in Refs. [\citen{Deotare:APL_121106:2009,Quan:APL_203102:2010,Quan:OE_18529:2011}]. The periodic mirror holes are chosen to maximise the Bloch band-gap, while the taper section consists of air holes linearly modulated down from the periodic mirror hole while keeping the filling factor constant. For example, three periodic mirror holes are inserted between two taper sections shown in Fig.~\ref{fig:strsim}(a), and only one hole in the tapered mirror shown in Figs.~\ref{fig:strsim}(b,c).
\\

\noindent{\bf{Optical and thermal properties simulations}}

\noindent{The nanobeam cavity supports a quasi-TM optical resonance and the calculated first order optical cavity mode profile is shown in Fig.~\ref{fig:strsim}(b).} Most of the $E_y$ field is confined between the two tapered mirrors, while the resonance is still strongly coupled to the waveguide mode indicating high on-resonance transmission. The cavity formed with one additional hole between the tapered mirrors has a calculated quality factor of $14900$ and effective mode volume of $0.8(\lambda/n)^3$ at wavelength $1557.83$ nm.

Using this calculated resonant spatial field profile as the heat source defined by $q_v(\mathbf{r})=Im[\epsilon(\mathbf{r})]|\mathbf{E(\mathbf{r})}|^2$, we investigate numerically the thermal behaviour of the nanobeam cavity by turning the heat source on at time $t=0$ and waiting for the structure to reach thermal equilibrium, then turning the heat source off and letting the structure cool. The temperature profile of the quasi-stationary thermal state at $z=0$ plane, under heat pulse shut down at $t=10$ $\mu$s, is shown in Fig.~\ref{fig:strsim}(c) indicating that the heat distribution is highly co-localised with the optical resonance field inside the small area in the cavity. 
By examining the temperature evolution at point ($0$, $0$, $0$) we determine that the thermal relaxation has the characteristic time of $780$ ns and the effective heat capacity of the nanobeam is $6.1\times10^{-12}$ J$\cdot$K$^{-1}$.\\

\begin{figure}[h!]
  	\includegraphics[width=0.98\columnwidth]{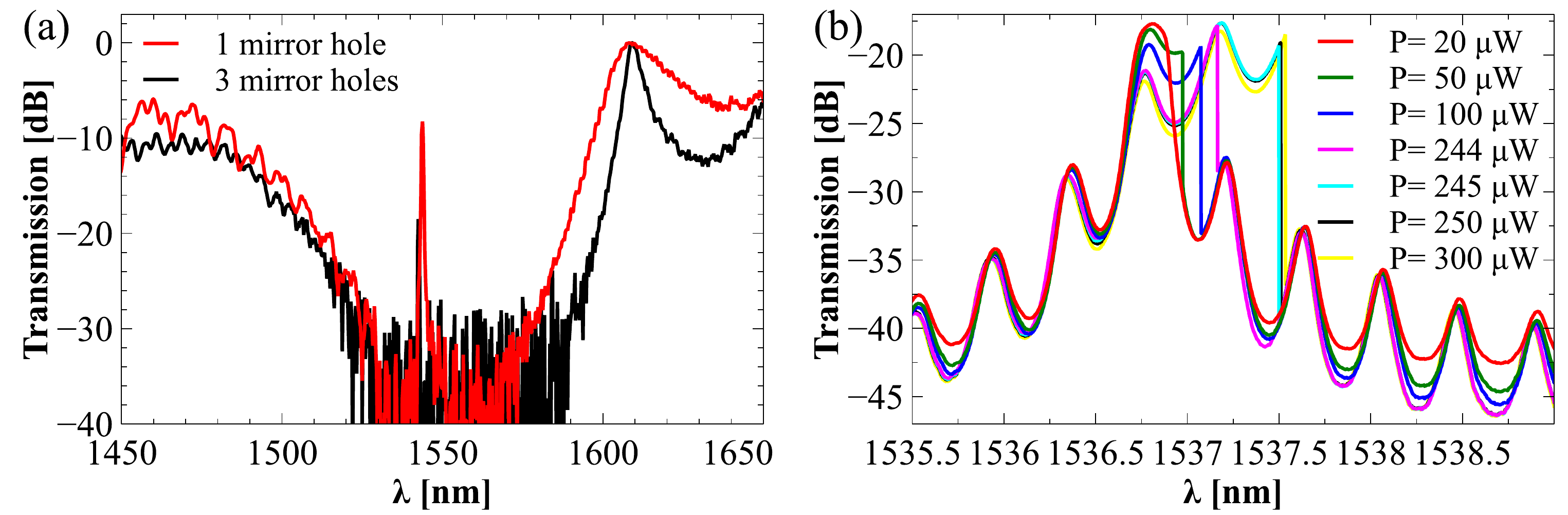}
	\caption{Experiment results of optical transmission. (a)~Transmission spectrum of the nanobeam cavity normalised to maximum transmission around $1610$~nm, measured using a broadband light source. (b)~Transmission in the vicinity of the resonance in sample with 1 mirror hole, as shown in Figs.~\ref{fig:strsim}(b,c), for different input powers, measured using a tunable laser source swept from short to long wavelengths.}
	\label{fig:exp1}
\end{figure}

\noindent{\bf{Experimental measurements}}

\noindent{We next experimentally characterise the thermo-optic property of the fabricated nano-beam cavity.} First, the cavity transmission was measured using a broadband supercontinuum source with sufficiently low power spectral density to avoid nonlinear thermo-optic effects. The transmission spectra of cavities with one and three mirror holes are shown in Fig.~\ref{fig:exp1}(a). Each of these curves has been normalised so the maximum transmission is $0$ dB. These curves indicate that the cavity resonance lies in a band-gap of approximately $50$ nm width and $30$ dB depth spanning from $1530$nm to $1580$nm, consistent with our design. Once the cavity resonances were identified, a tunable laser was used to probe the sample transmission in the vicinity of the resonance peak to obtain more accurate spectral data in the linear regime. With a coupled laser power of $1$ $\mu$W, we find that the first order resonance of the nanobeam cavity with one mirror hole has a quality factor of $10^4$ at a wavelength of $1542.99$ nm. The cavity with three mirror holes has a quality factor of $~3\times10^4$ at a wavelength $1541.85$ nm, but this comes at the cost of lower on-resonance transmission, as can be seen in Fig.~\ref{fig:exp1}(a).

To study the transmission spectrum of the cavity with one mirror hole at different input powers we use a continuous-wave tunable laser to sweep across the resonance peak 
from short to long wavelengths at a speed of $500$ pm/s with the step size of $1$ pm. The transmission spectra start to show asymmetry at input powers of $20$~$\mu$W, exhibiting a sudden drop of approximately $10$ dB at a wavelength just beyond the resonance peak, as shown in Fig.~\ref{fig:exp1}(b). We refer this wavelength as critical switch-off wavelength from here onwards. We note here that the experimental optical powers specified throughout this paper are the estimate of the powers in the nanobeam, taking into account a $10$ dB coupling loss.

With increasing input power the asymmetry becomes clearer and the critical switch-off wavelength shifts to longer wavelengths, as shown in Fig.~\ref{fig:exp1}(b). The transmission linewidth broadening is attributed to the thermo-optic nonlinearity~\cite{Brissinger:PRB_033103:2009}. In addition to the linewidth broadening we also observe the periodic fringes imprinted on the Lorentzian-shaped resonance peak. It has previously been reported that such additional spectral features are introduced by the Fabry-P\'{e}rot cavity formed by the cleaved facets of the access waveguide~\cite{Velha:APL_171121:2006, Brissinger:PRB_033103:2009}. The oscillations are observed in our sample at all input powers above the threshold power $20$~$\mu$W even though we intentionally angle the waveguides at $7$ degrees off-normal to the edges of the chip in order to reduce reflection at the cleaved facets.

An interesting feature appears in Fig.~\ref{fig:exp1}(b) when the input power increases from $244$~$\mu$W (magenta line) to $245$~$\mu$W (cyan line): the critical point where the transmission drop occurs, shifts discontinuously to longer wavelengths by $0.7$~nm. The $1$~$\mu$W power step is limited by the experiment setup, so it is not possible to resolve this sudden shift more accurately.  After this sudden wavelength jump, the critical wavelength shifts with increasing power at a much slower rate than before the jump. In the following section we demonstrate that this discontinuous critical point results from the isolated thermal equilibrium induced by the coupling between Fabry-P\'{e}rot and nano-beam cavities.

\begin{figure}[h!]
   	\includegraphics[width=0.9\columnwidth]{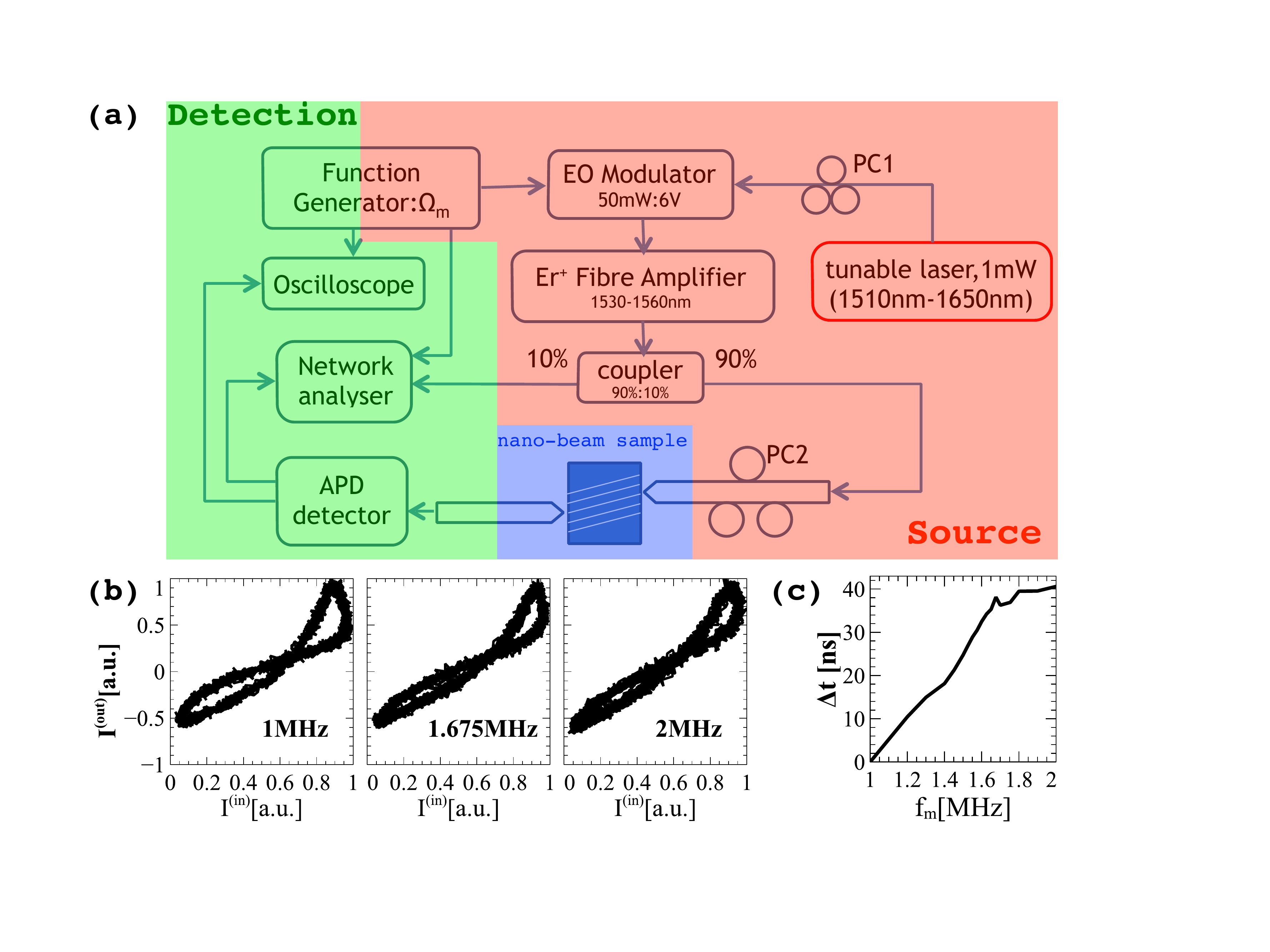}
	\caption{(a)~Experiment setup used to measure the dynamic thermo-optic response to sinusoidal modulated signals. (b)~Transmitted vs. input optical signal (Lissajous curves), at modulation frequency $1$ MHz, $1.675$ MHz and $2$ MHz from left to right for $360\mu$W average input power. (c)~Time delay of output signal with respect to input signal as a function of modulation frequency at $360\mu$W average input power.}
	\label{fig:exp2}
\end{figure}

To investigate the nonlinear dynamic characteristics of the thermo-optic response, the sample was excited by a modulated continuous wave source using the experimental arrangement shown in Fig.~\ref{fig:exp2}(a). The amplitude modulation was driven by a single frequency RF signal at frequency $f_m\in$ ($1$~MHz, $2$~MHz). The average power entering the lensed fibre at PC2 was kept constant at $360\pm 3\mu$W for all driving frequencies and the tunable laser was set to the wavelength which gives the maximum nonlinear transmission amplitude.

At such a large thermal offset which corresponds to high temperature rise, the nonlinear response of the cavity generates second and third harmonics of the modulated input signal. These nonlinear contributions can be visualised using Lissajous curves in which the output signal is plotted against the input signal as in Fig.~\ref{fig:exp2}(b). Lissajous curves are often used to distinguish complex harmonic signal content and their shape is very sensitive to the frequency ratio of two signals. In our measurement, we observe open circles and asymmetric lobes in the Lissajous curves which indicates that the output signal has strong second and third harmonic components, respectively. The amplitude of the second harmonic component reaches its maximum at $1$ MHz modulation frequency and decreases monotonically as the frequency increases to $2$ MHz. This is evident by the closing of the asymmetric lobes with increasing modulation frequency.

The suppression of the second harmonic component occurs when the characteristic time for heat diffusion becomes too slow to follow the optically induced heating. According to the numerical simulations, the thermal decay time of the suspended cavity is $780$ ns (this is the time it takes for the temperature to reduce to $1/e$ of the maximum value once the heat source is turned off). Thus, when the optical signal that creates the local heat source is modulated close to and above a frequency of $1.28$ MHz ($=1/780$ ns) the thermally-induced second harmonic response is suppressed.

We also measure the time delay between the input and output signals, see Fig.~\ref{fig:exp2}(c). The time delay is extracted by finding the time offset that gives the maximum overlap integral of input and output signals, where we define the time delay at $1$ MHz to be zero as a reference. This analysis shows that faster modulation results in longer time delay at frequencies where the thermal relaxation can follow the optical heating modulation. Within the frequency range corresponding to the characteristic time, the optical energy absorbed into the cavity within a certain time increases linearly with the input modulation frequency when the average power per period is constant, therefore, the time delay increases linearly with the frequency. Beyond this frequency range, such as modulation frequency between $1.65$ MHz and $2$ MHz, the time delay saturates as the heat transfer cannot follow the increasing heat absorption.

\section*{Discussion}

The experimental results presented above exhibit somewhat different characteristics to previous reports of thermo-optic nonlinearities in photonic nanocavities~\cite{Velha:APL_171121:2006, Brissinger:PRB_033103:2009}. We discuss and explain the main differences in this section.

\begin{figure}[h!]
   	\includegraphics[width=0.98\columnwidth]{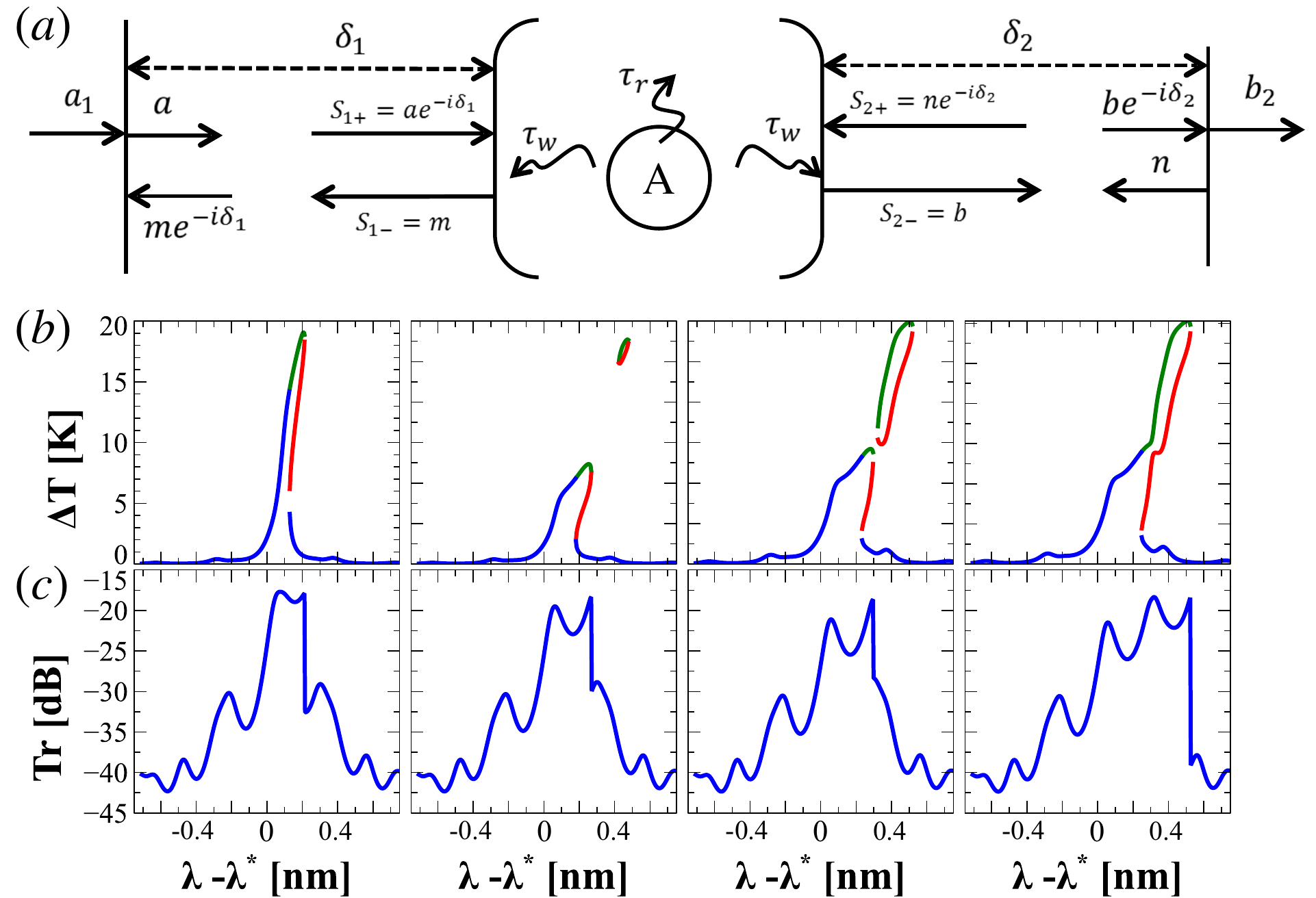}
	\caption{Theoretical modelling. (a)~Schematic representing the coupled mode model describing coupling between the Fabry-P\'{e}rot cavity formed by the waveguide end facets, and the nanobeam cavity. (b)~Cavity temperature increase predicted by Eq.\ref{eq:thermal} at input power $0.5$ mW, $1.5$ mW, $3.0$ mW and $3.5$ mW respectively, from left to right. (c)~Modelled transmission corresponding to the thermal equilibria shown in (b), as a function of input wavelength. We choose $\lambda^*=1535$ nm as the reference wavelength.}
	\label{fig:theory}
\end{figure}

The periodic fringes that overlay the cavity resonance transmission spectra in Fig.~\ref{fig:exp1}(b) result from coupling between the nanobeam cavity and a parasitic Fabry-P\'{e}rot (FP) cavity which is formed by the cleaved end facets of the access waveguide. This is evident from the $0.8$~nm fringe spacing in the transmission spectra, which is consistent with the $\sim 3$ mm optical length of the cleaved sample. The coupling between this FP cavity and the nanobeam cavity can be described by a model in which the nanobeam cavity resonance of amplitude $A$ is treated as a point inside a FP cavity of single trip phase shift $\delta_1+\delta_2$, as shown in Fig.~\ref{fig:theory}(a). Following established coupled mode theory~\cite{Joannopoulos:2008} and the transmission line method, the various field amplitudes in the locations indicated in Fig.~\ref{fig:theory}(a) can be related as

\begin{subequations}\label{eq:FPcavity}\begin{align}
\begin{pmatrix}
  m \\
  b
 \end{pmatrix}
&=
\begin{pmatrix}
 \mathfrak{T}-1 &  \mathfrak{T}\\
 \mathfrak{T} &  \mathfrak{T}-1
 \end{pmatrix}
\begin{pmatrix}
 ae^{-i\delta_1} \\
 ne^{-i\delta_2}
 \end{pmatrix},\\
a &= jta_1 + rme^{-i\delta_1},\\
n &= rbe^{-i\delta_2},\\
b_2 &= jtbe^{-i\delta_2},
\end{align}
\end{subequations}
where $t$ and $r$ are respectively the transmission and reflection ratios at the chalcogenide glass - air interface, $\omega_r$ is the resonant angular frequency of the nanobeam cavity, $\tau_w$($\tau_r$) is the dissipation rate from the nanobeam cavity to the access waveguide (far-field radiation channels), and $\mathfrak{T}= (2/\tau_w)/[j(\omega-\omega_r) + 1/\tau_r + 2/\tau_w]$ stands for the transmission ($S_{12}$) parameter of the nanobeam cavity~\cite{Joannopoulos:2008}.

Using this framework, the amplitude inside the nanobeam cavity, $A$, can be expressed as a function of $\mathfrak{T}$, $f(\mathfrak{T})$ as 

\begin{equation}\label{eq:NBcavity}
A=f(\mathfrak{T})=\frac{\mathfrak{T}\tau_w}{2} \frac{jte^{-i\delta_1}(1+re^{-i2\delta_2})}{C_1C_2-C_3^2} a_1,
\end{equation}
where $C_1 = 1 - r(\mathfrak{T}-1)e^{-i2\delta_1}$,  $C_2 = 1 - r(\mathfrak{T}-1)e^{-i2\delta_2}$, and $C_3 = r\mathfrak{T}e^{-i(\delta_1+\delta_2)}$.

According to the numerical thermal simulation, the normalised temperature profile during relaxation $\Delta{T(\mathbf{r},t)}$ remains self-similar and close to the quasi-stationary distribution $\Delta{T(\mathbf{r},t_s)}$, where $t_s$ is the time when the quasi-stationary state appears. It is evident by the fact that their normalised overlap over the chalcogenide glass beam is close to unity, $\int\Delta{T(\mathbf{r}, t_s)}\Delta{ T(\mathbf{r},t)}dV/\int|\Delta{T(\mathbf{r}, t_s)}|^2dV \ge 0.957$ at all times. Thus we consider the heat diffusion in the time domain and write the thermo-optic coupling equations as
\begin{subequations}\label{eq:thermal}
\begin{align}
\Delta\dot{T}(t) &= -\frac{1}{\tau} \Delta T(t) + \frac{\eta}{C_p}\frac{\omega}{2\pi}|A(t)|^2,\\
A&=f(\mathfrak{T}(\omega_r=\omega_o+C_t\Delta{T})).
\end{align}
\end{subequations}
Here $\tau$ is the characteristic time of the heat diffusion in the suspended chalcogenide glass nanobeam cavity, $\eta$ is the heat absorption ratio, $C_p$ is the effective heat capacity, $C_t$ is the rate of change of the resonant frequency due to temperature change, and $\omega_o$($\omega_r$) is the resonant angular frequency at room temperature $T_0 = 293.15$~K (temperature $T_0 + \Delta{T}$).

According to the theoretical analysis, the FP cavity modifies the thermal equilibrium via the optical field amplitude modulation inside the nanobeam cavity, strongly affecting the nonlinear transmission behaviour. Fig.~\ref{fig:theory}(b) shows the modified equilibrium temperature of the cavity for different laser input powers, and the corresponding transmission spectra are presented in Fig.~\ref{fig:theory}(c). The curves were calculated for parameters obtained in simulations: $\tau=780$ ns, $C_p=6.1\times10^{-12}$ J$\cdot$K$^{-1}$, in literature: $C_t= -6.1294\times10^9$ rad$\cdot$K$^{-1}$~\cite{RPWang:2014}, and in fittings to experimental results Fig.~\ref{fig:exp1}: $\lambda_o=1535.06$nm, $\eta=2.7\times10^{-4}$, $r=0.12$, $t=0.1176$, $l_1=1.3$ mm, $l_2=1.7$ mm, $\tau_r=9.59\times10^{8}$ Hz and $\tau_w=3.84\times10^{10}$ Hz.

The simulation predicts the discontinuous switch-off wavelength observed in the experiment, as can be seen by the jump between transmissions at input power $3.0$ mW and $3.5$ mW in Fig.~\ref{fig:theory}(c), associated with the appearance of an isolated thermal equilibrium island. Figure~\ref{fig:theory}(b) shows the co-existing temperature equilibrium states (coloured in blue, green and red, respectively) which are responsible for the thermo-optic bistability enabled at the input power $0.5$ mW. Further increasing input power to the range from $1.5$ mW to $3.5$ mW separates the bistable wavelength range until the isolated thermal equilibrium island grows large enough to connect to the accessible thermal equilibria. The connection results in the switch-off wavelength jump since the accessible equilibrium extends abruptly to the previously inaccessible state.

\section*{Conclusion}

{\noindent}To conclude, we have investigated nonlinear thermo-optic effects in suspended chalcogenide glass nanobeam cavities, and identified a previously unexplored nonlinear effect due to coupling of the nanobeam cavity and Fabry-P\'{e}rot resonances of the whole waveguide. Nanobeam cavities with optical quality factor $10^4$ around $1550$ nm, thermal characteristic time $780$ ns and effective heat capacity $6.1\times10^{-12}$ J$\cdot$K$^{-1}$ were designed and fabricated using Ge$_{11.5}$As$_{24}$Se$_{64.5}$ glass. Thermo-optic bistability was demonstrated by the asymmetric and discontinuous transmission spectrum measured when the excitation wavelength was swept across the resonance peak from short to long wavelengths. We also observed a discontinuous shift of the critical wavelength with increasing input power. We have developed a theoretical model to explain this behaviour and shown that it is due to an isolated thermal equilibrium state arising from coupling to a parasitic Fabry-P\'{e}rot cavity. Finally, we used a variable-frequency modulated input source to investigate experimentally the nonlinear thermo-optic effect by measuring the cavity temporal response at high temperature rise. Modulation frequencies faster than the characteristic thermal response time of the cavity were found to exhibit saturated time delays and suppressed generation of the second harmonics of the modulated input signals.

This analysis reveals new insights into thermal nonlinearities in optical nanocavities, and the surprising role of parasitic resonances induced by reflections from waveguide facets. Despite being relatively weak, these reflections can have a significant impact on the nonlinear response of the cavity, resulting in additional equilibrium states and unexpected transitions between them.  Understanding and controlling such effects can be essential to exploit switching in nanophotonic devices.

\section*{Methods}

\noindent{\bf{Optical Cavity Mode Simulation}}

{\noindent}The first order optical resonance of the nanocavity is calculated using three-dimensional full-vector simulations in MEEP, a free finite-difference time-domain (FDTD) simulation software package~\cite{Oskooi:2010-687:ComputerPhysicsComm}. The resonant frequency and quality factor are extracted from the time evolution of the $E_y$ response at point ($58$ nm, $0$ nm, $0$ nm) by Harminv~\cite{Mandelshtam:1997-6756:JChemPhys}, a free program to solve the problem of harmonic inversion, after the point Gaussian current source is turned off. The $E_y$ cavity mode profile is obtained in a second FDTD simulation using a narrow bandwidth gaussian source centred at the cavity resonant frequency $300$ time steps after the excitation is turned off. Throughout the simulation, the refractive index of chalcogenide glass and silica are set to $2.63$ and $1.45$, respectively.
\\

\noindent{\bf{Thermal Properties Simulation}}

{\noindent}The thermal properties of the nanobeam are calculated using time-dependent simulations with COMSOL {\it heat transfer in solids} module. The temperatures of the outer boundaries of air and silica far from the cavity are assumed to be at room temperature $293.15$ K. The source of $Im[\epsilon(\mathbf{r})]|\mathbf{E(\mathbf{r})}|^2$ is applied to the Chalcogenide beam cavity for $10$ ms to heat up the structure to a stable state before shutting off the source to allow the nanobeam to cool down. The density, heat capacity and thermal conductivity of the chalcogenide glass and silica are set to ($ 4490 [kg\cdot m^{-3}]$, $300 [J\cdot kg^{-1}\cdot K^{-1}]$, $0.9 [W\cdot m^{-1}\cdot K^{-1}]$) and ( $2651 [kg\cdot m^{-3}]$, $703 [J\cdot kg^{-1}\cdot K^{-1}]$, $1.38 [W\cdot m^{-1}\cdot K^{-1}]$), respectively~\cite{RPWang:2014, Adam:2013, Gan:2014-377:JRamanSpectrosc, Weber:2002}.  In the simulations, the impact of air convection is neglected and the thermal properties of air are set as functions of the temperature to describe the heat diffusion as provided by COMSOL material library.
\\

\noindent{\bf{Device Fabrication}}

{\noindent}The device is patterned in ZEP-520A resist by e-beam lithography. The nanobeam cavities are exposed using area exposure mode operating at $30$ kV gun power and $7.5$ $\mu m$ aperture, and the access waveguides are exposed using fixed beam moving stage (FBMS) mode and $20$ $\mu m$ aperture. The access waveguide is written in the same exposure that defines the cavity pattern to minimise the coupling loss due to misalignment.The writing fields of the two apertures are aligned through iterative adjustment of the writing field alignment.

After exposure and development the resist pattern is transferred to the chalcogenide membrane by Inductively Coupled Plasma (ICP) etching using CHF$_3$ gas. The nanobeam cavity is then released from the substrate by removing the 100nm thick SU-8 sacrificial layer using Oxygen dry etching. Finally, the device is passivated with a $2$ nm Al$_2$O$_3$ film deposited by atomic layer deposition (ALD) to protect the chalcogenide glass from further crystallization.
\\

\noindent{\bf{Optical Measurement Setup}}

{\noindent}The optical signal is coupled into/out-of the nanobeam cavity through the cleaved end facets of access waveguides using lensed fibres. The transmission spectrum, shown in Fig.~\ref{fig:exp1}(a), was characterised using a super-continuum light source with total optical power $\sim 200$~mW in the wavelength range from $600$~nm to $1700$~nm. A polariser and an in-line fibre polarisation controller were used to align the input light polarisation to the nanobeam cavity to maximise the output intensity in $TM$ mode ($H_z$ field dominates). The transmission spectrum was measured with an optical spectrum analyser (ANDO AQ6317B) using $0.1$ nm resolution and 'HIGH 1' sensitivity.
For the dynamic thermo-optic response measurements, a high resolution ($1$ pm) tunable laser (JDS FITEL $SWS15101$) set to $1$ mW output power was used for all modulation frequencies. The linearly polarised input light signal was sent to a polarisation controller (Agilent $11896$) PC1 before being modulated using an electric-optic (EO) modulator (JDS Uniphase $10$~Gb/sec "Bias ready" intensity modulator) driven by a function generator (SONY-Tektronix $AWG2040$ arbitrary waveform generator), see Fig.~\ref{fig:exp2}(a). The input light polarisation, and driving conditions of the modulator were adjusted to achieve maximum modulation depth. The modulated signal was then amplified with an erbium doped fibre amplifier (Amonics AEDFA-LP) then split into two parts: $10\%$ of the power was sent to a network analyser (Agilent infiniium DCA $86100A$ wide-bandwidth oscilloscope) to monitor the input power; the remaining $90\%$ of the power was sent to the sample via an in-line fibre polarisation controller PC2. The transmitted output light was collected in a second tapered fibre, and detected using a AC-coupled avalanche photo-diode (APD) detector (Lab buddy $R402APD$). The detected electric signal was sent to the network analyser for comparison to the input signal. Note here that the APD detector is used because it has a better power resolution of around $200$~nW, $100$ times smaller than that of the network analyser.



\section*{Acknowledgements}

This work was supported by the Australian Research Council (ARC) Discovery Project DP130100086. Numerical simulations were performed with the assistance of resources provided at the NCI National Facility systems at the Australian National University supported by the Australian Government. E-beam lithography was performed in part at the ACT node of the Australian National Fabrication Facility, a company established under the National Collaborative Research Infrastructure Strategy to provide nano and micro-fabrication facilities for Australian researchers.

\section*{Author contributions statement}
Y.S., T.P.W. and D.Y.C. carried out experiments, Y.S. and A.A.S. performed theoretical analysis. All authors conceived the work, planned experiments, analysed results, wrote and approved the final manuscript.

\section*{Competing interests}
The authors declare no competing financial interests.

\section*{Corresponding author}
Correspondence to Yue Sun.

%
%
%

%
%
%
\end{document}